\documentclass[sigconf, nonacm]{acmart}

\usepackage{multirow}
\usepackage{multicol}
\usepackage[inline]{enumitem}
\usepackage{hyphenat}
\usepackage{float} 

\newcommand\leftHeaderTableWidth{1.2in} 

\begin{document}
\title{A survey of open-source data quality tools: shedding light on the materialization of data quality dimensions in practice}

\author{Vasileios Papastergios}
\affiliation{
    \institution{Aristotle University}
    \state{Greece}
}
\email{papastva@csd.auth.gr}

\author{Anastasios Gounaris}
\affiliation{
    \institution{Aristotle University}
    \city{Thessaloniki}
    \state{Greece}
}
\email{gounaria@csd.auth.gr}

\begin{abstract}
    Data Quality (DQ) describes the degree to which data characteristics meet requirements and are fit for use by humans and/or systems. There are several aspects in which DQ can be measured, called DQ dimensions (i.e. accuracy, completeness, consistency, etc.), also referred to as characteristics in literature. ISO/IEC 25012 Standard defines a data quality model with fifteen such dimensions, setting the requirements a data product should meet.
    In this short report, we aim to bridge the gap between lower-level functionalities offered by DQ tools and higher-level dimensions in a systematic manner, revealing the many-to-many relationships between them. To this end, we examine 6 open-source DQ tools and we emphasize on providing a mapping between the functionalities they offer and the DQ dimensions, as defined by the ISO standard. Wherever applicable, we also provide insights into the software engineering details that tools leverage, in order to address DQ challenges.
\end{abstract}

\maketitle

\section{Introduction\label{section:Introduction}}

In the modern data-driven era, the importance of Data Quality (DQ) has escalated, as organizations increasingly rely on data to drive strategic decisions and operations. \emph{High-quality} data are essential for ensuring the integrity of analytics, empowering Machine Learning (ML) models and supporting Business Intelligence (BI) efforts. Poor data quality can lead to costly, misguided decisions, operational inefficiencies, and a loss of trust among stakeholders or end users. Managing DQ effectively becomes even more crucial as the volume of data collected, stored and processed grows exponentially.

In response to these challenges, a wide variety of DQ tools have been developed, providing solutions that help organizations enhance their DQ capabilities. However, despite the great amount of effort devoted to address DQ challenges, there still exist significant discrepancies in (i) the terminology used to describe these solutions and (ii) the alleged connection of these solutions with the DQ dimensions.

In particular, there are functionalities that are common among the tools; yet different names are employed to describe them. For example, ensuring that all records of a dataset satisfy some kind of restriction can be found as ``conformance", ``compliance'' or even ``validity''; all referring to the same software engineering functionality. In addition, there are cases where the same term can be interpreted to different functionality in different tools. ``Completeness'', for  example, may refer to the number of records in a dataset in some tools, while associated with the presence of NULL values in others. Even the DQ dimensions are not always named under the same terms. For instance, the words ``currentness'', ``freshness'', ``recency'' and ``timeliness'' can all be found in literature and tools' documentation, referring to roughly the same DQ dimension.

To date, there has not been much work conducted in terms of connecting the low-level functionalities provided by the tools with the dimensions they are closely related to. In simple words, it seems that a bridge is missing between the dimensions defined by some theoretical DQ model and the way real-world implementations materialize them taking into consideration that lower-level functionalities are related to higher-level dimensions in a many-to-many manner.

In this paper, we aim to bridge this gap. We investigate six widely-spread open-source DQ tools. We opt for using ISO/IEC 25012 Standard~\cite{iso25012} as the unifying theoretical basis of our investigation. This standard defines a DQ model with 15 dimensions (characteristics), i.e,  requirements that a data product has to meet. The quality characteristics in the standard are shown in Figure~\ref{img:iso25012}. Overall, our work seeks to offer a concise overview of the current landscape in DQ management, as represented by the  investigated tools. In particular, our main contribution is twofold and can be summarized in the following:

\begin{figure}[tb!]
    \centering
    \includegraphics[width=0.8\columnwidth]{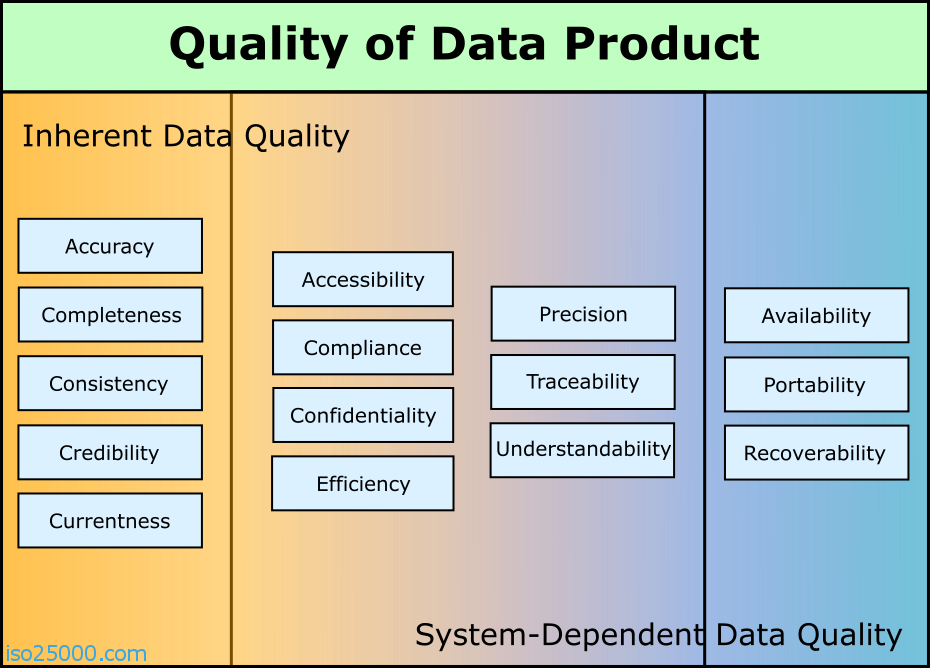}
    \caption{The data quality characteristics as defined in ISO/IEC 25012 ~\cite{iso25012}.}
    \label{img:iso25012}
\end{figure}

\begin{enumerate}
    \item  We provide a comprehensive and unifying listing of all the different functionalities offered under different names and in multiple variations in 6 widely used open-source DQ tools. Wherever applicable, we also dive into engineering insights regarding the implementation of these functionalities.

    \item Taking a step further, we focus on the connection between existing solutions and DQ dimensions. More specifically, we present a mapping between the DQ low-level functionalities observed in the tools and the standard's dimensions that are closely associated with them.
\end{enumerate}

The rest of our work is structured as follows. In Section~\ref{section:methodology}, we summarize the methodology.
In Section~\ref{section:survey_findings}, we present the findings of our investigation and introduce the mapping between DQ solutions and dimensions. Section~\ref{section:related_work} provides information about related work and how it compares with ours. Finally, in Section~\ref{section:conclusion_future_work}, we conclude providing also insights into future work.

\section{Our Methodology\label{section:methodology}}
In this section, we explain our approach in terms of conducting our survey. We begin with selecting representative open-source DQ tools in line with other existing selection, e.g., \cite{atlan_list}, enriched with tools we are aware of, such as Apache Griffin~\cite{griffin_website}. The final list comprises 6 widely-used, open-source and \emph{active} projects. To the best of our knowledge, there are not other tools that meet these specifications and are dedicated to DQ management. The tools are:
\begin{enumerate*}
    \item Deequ,
    \item dbt Core,
    \item MobyDQ,
    \item Great Expectations (GX),
    \item Soda Core and
    \item Apache Griffin
\end{enumerate*}.

The second step of our investigation has been to emphasize on the low-level functionalities offered by each tool. The challenging part of this step has been to bypass the widely diverse terminology and focus on the core functionality, grouping together similar functionalities from different tools. To accomplish this, we have dived into the source code of these tools, in combination with studying their official documentation, setting them up locally on our machines and experimenting with simple examples.

Based on the recorded functionalities, we, then, proceed to revealing the connections between the recorded functionalities and the DQ dimensions, as defined by the ISO/IEC 25012 standard. The utter goal of such an approach has been to provide useful insights into the way DQ tools available at current time address and materialize these dimensions.

\section{DQ functionalities and mappings\label{section:survey_findings}}

The findings of our investigation are summarized in Table~\ref{table_functionalities_to_dimensions}. The second column contains all the different low-level functionalities (i.e., data checks) that were found in the tools examined. The functionalities are grouped into more general categories, listed in the first column. The next six columns represent the ISO/IEC 25012 dimensions that are affected by at least one low-level functionality. Note that, out of the fifteen dimensions defined by the standard, we were able to identify six of them as associated with the recorded functionalities, the ones referring to the so-called inherent data quality. We use a $\bullet$ to denote that a low-level functionality is closely related to a specific dimension. The rest six columns stand for the examined DQ tools. For each recorded functionality, we use a $\checkmark$ to imply that it is \emph{directly} supported by a specific tool. Note that we only mark the respective cell if the low-level functionality is \emph{squarely} supported by some built-in check of the tool. This does not necessarily imply that the functionality cannot be achieved by other tools, as well, in an indirect or user-defined way. In the subsections that follow, we provide details about the functionality categories, shifting our emphasis towards their connection with the DQ dimensions.

We note that a comprehensive list of the \emph{exact names} of the functions provided by the tools for each low-level functionality can be found in Table~\ref{table_functionalities_names} in the appendix. Some names are hyphenated to fit in the line width. In these cases, the actual name does not contain a hyphen.

\begin{table*}[ht]
    \scriptsize
    \begin{center}
        \bgroup
        \def\arraystretch{2}
\begin{tabular}{|c|c|c|c|c|c|c|c||c|c|c|c|c|c|}
            \cline{3-14}
            \multicolumn{1}{l}{}                                              & \multicolumn{1}{l|}{}                                                                                      & \multicolumn{6}{c||}{DQ dimensions}                                       & \multicolumn{6}{c|}{DQ tools}                                                                                                                                                                                                                                                                                                                                                                                                                                                                                                                                                                                                                                                                                                                                                                                                                              \\
\end{tabular}
        \egroup
    \end{center}
    \caption{A catalog of the low-level functionalities (column 2) offered by the examined DQ tools, grouped into general categories (column 1). Each low-level functionality is associated with the dimension(s) they are closely related to (columns 3-8) and the tools they are directly supported by (columns 9-14).}
    \label{table_functionalities_to_dimensions}
\end{table*}

{\bf Anomaly Detection.}
Anomaly detection (AD) is closely related to the accuracy dimension. An anomaly is an erroneous and/or unexpected value that does not represent the truth (actual value) of the intended attribute; as such, it violates the accuracy definition provided by the standard. AD functionality is offered by Deequ and is based on previously saved metrics for the same or similar data, performing comparisons with some kind of user-defined threshold(s). The tool has a wide range of built-in (heuristic, non-ML) anomaly detection strategies. The strategies cover both absolute and relative metrics, and offer support for batch (mainly) and streaming data. All AD strategies can be combined with previously saved metrics. For example, given a saved metric about the number of rows in a dataset, an AD check could be to examine whether the saved size and the currently computed size of the dataset exhibit a difference that falls inside a range (either in absolute or relative value).

    {\bf Statistic Measures.}
The computation of statistic measures (data profiling) on single columns is a common low-level functionality found in the majority of the examined tools. Minimum and maximum values, sum, standard deviation (std), etc. are examples of such measures. Deequ, GX and Soda Core directly offer these checks as built-in functionalities, while the rest of the tools offer indirect ways to the user to define their own checks that fall into this category. In the general case, the user defines a threshold and a comparison operator. The tool is, then, responsible for executing the test. The general formula that can be derived is the following:

\begin{displaymath}
    m \text{\hspace{1mm}} \{>, <, \ge, \le, = , \neq\} \text{\hspace{1mm}} T,
\end{displaymath}

where $m$ is a computed statistic measure and $T$ a user-defined numeric threshold. The check outcome is binary, i.e. pass or fail.

These low-level functionalities are closely related to three DQ dimensions: accuracy, consistency and compliance. In particular, accuracy is strongly connected, since statistic measures usually encapsulate the true value of the intended attribute. For example, a column representing product quantities is expected to have non-negative ($min = 0$) values, or else data accuracy is impaired. Consistency is also connected, because measures such as mean, median and std take into account not only a single value, but also the coherence of it with other data as well in a specific context. Compliance is also relevant, since the use of thresholds for computed measures implies the presence of (physical or business) rules the data have to comply with.

From an implementation perspective, all tools leverage SQL-like queries, which are executed at the data sources to obtain the desired statistic values. An interesting implementation detail comes from Soda Core and it has to do with the way standard deviation and variance are computed. There are two variations provided, which use either the complete dataset or just a sample of it, respectively. Under the hood, again, the tool leverages the \texttt{pop} and \texttt{samp} functionality provided by SQL, in order to run queries on the data sources.

    {\bf Column-level validity.}
Another family of checks provide low-level functionalities for validating column values in a column-level scope. Several variations of these checks were found throughout our investigation. In particular, dbt Core offers a check that examines whether values in a column are not all the same. Given an examined column with values, this check can be summarized in the following clause:

\begin{displaymath}
    \exists \text{\hspace{1 mm}} v(i) \neq v(j), \text{\hspace{2 mm}} i, j \in [0, N),
\end{displaymath}

where $v(i)$ and $v(j)$ denote two values in the column at index $i$ and $j$ respectively and $N$ denotes the total number of values in the examined column. The outcome of the check is binary (pass or fail). Other variations of this check, provided by dbt Core and GX, take it a step further by examining whether the values have a specific ordering, rather than being just different. More specifically, these checks require that the values are sequential, increasing or decreasing. The math formula for this check can be summarized in the following

\begin{displaymath}
    v(i) \text{\hspace{1 mm}} \{>, <, \ge, \le\} \text{\hspace{1 mm}} v(i-1), \text{\hspace{2 mm}} \forall \text{\hspace{1 mm}} i \in (0, N)
\end{displaymath}

where $v(i)$ and $v(i-1)$ are two consecutive values in the column, at indices $i$ and $i-1$ respectively and $N$ is the total number of values in the examined column. The outcome of the check is, again, binary.

Accuracy is closely related to this functionality because, in case the check fails, it can be assumed that the values stored in the data are not the true ones. Consistency is also affected, since values violating ordering or inequality are not coherent with the rest of the data, in a specific context of use. Furthermore, there must be some convention, enforcing that the values have to be different. As a result, this category of low-level functionalities is closely connected with compliance, as well. Regarding implementation, all the above are executed in some form of SQL queries with small differences in each tool.

    {\bf Distribution / Quantiles.}
Concerning the values of a specific column, Deequ offers built-in functionality that enables the user to query on the distribution of the column values. More specifically, the tool constructs a distribution object with a single pass on the data. The latter stores distribution information that can be accessed and checked by the user, using complex constraints. The supported functionalities by Deequ include both absolute and ratio checks on the distribution values. Regarding implementation, callback functions can be passed as arguments by the user. These functions can  be called either on every element or on an aggregated value of the distribution object. The affected dimensions are accuracy, consistency and compliance for the same reasons as above.

Similarly, there are checks that measure the distribution quantiles of a column's values. Deequ provides three flavors of the quantile distribution check. The first flavor computes exact percentiles. The second flavor~\cite{approximate_quantile_deequ} computes an approximate quantile, based on the Apache Spark's \texttt{sql.catalyst} relative implementation. The third flavor leverages KLL Sketching~\cite{kll_sketch}. After quantile computation, the user can define and execute threshold-based checks. Nevertheless, the distribution of values in current time may not be valuable information alone, but only when compared with the distribution of the same attribute in the past. Soda Core comes with an interesting low-level functionality about the distribution change of a column's values, based on some saved distribution snapshot.

    {\bf Record-level validity\label{record_level_validity}.}
Apart from the value's distribution, several tools provide low-level functionalities to enable the user check that values in a column are limited to a specific set or range of allowed values. Accuracy is directly affected by this check, due to the fact that non-allowed values are not truthfully representing the modeled domain. Compliance is also affected, since a regulation is implied, imposing a set or range of allowed values.

Regarding software engineering implementation, Dbt Core offers support for both inclusive and exclusive ranges. The check can be formally expressed with the following:
\[ \forall i \in [0, N), v(i) \in \mathcal{S}_{allowed}, \]
where $N$ is the total number of values in the examined column, $v(i)$ is the value at index $i$ and $\mathcal{S}_{allowed}$ is the set (or range) of allowed values. It also supports a where-clause to execute the check only on a subset of the total records.
Other variations, found in GX, are (i) checking that the \emph{most common} value is contained in a user-defined set and (ii) involving two columns at a time, examining whether the pair of values are contained in an accepted set of pairs.

Being able to define in advance all different alternatives contained in an accepted range/set is not always the case, however. For that purpose, GX and Soda Core provide functionalities such as verifying length constraints on strings or digit restrictions on numbers. These checks can be considered a generalization of the previous cases, where the allowed values are all possible alternatives. From an engineering point of view, the tools opt for selecting violating records, using some form of SQL queries. Enforcing compliance with some SQL-predicate or regex are more generic forms these low-level functionalities can take. Note that checking regex using SQL is enabled through the \texttt{match\_recognize} operator.

    {\bf Cross-column checks.}
Up to that point, we have examined checks that operate on single columns or combinations of them. However, there are cases where the objective is to enforce some kind of constraint that involves more than one column, for a specific record of data. Comparisons between respective values of two different columns within the same record is an example of such functionality, offered by GX. The tool leverages SQL queries that capture records violating the check. The outcome is a set violating records ($\mathcal{S}_{viol}$) and can be formally expressed as following:
\[ \mathcal{S}_{viol} = \{ r_i : r_i.a < r_i.b \}, \]
where $r_i$ is a record in the dataset with attributes $a$ and $b$. The rule that is checked between the two values can be any comparator or, in general, any boolean callback function that takes as input the two values and outputs a boolean value.

This check is closely related to three dimensions, as previosuly: accuracy, because a violating record does not represent the true value; consistency, since a violating value is not coherent with the rest values; compliance, due to the presence of the validation rule. Other low-level functionalities in this category are checking restrictions on either the correlation or the mutual information between two different columns, with straightforward implementations.

    {\bf Cross-row checks.}
Another type of low-level DQ functionality relates to checks regarding multiple rows together. The check ensures that columns representing range information do not have illegally overlapping values, where ``illegally'' can be user defined. Note that the check requires the existence of two columns, one representing the lower and one the upper bound of the range. The tool sorts the data based on the lower bound column and, then, leverages the \texttt{LEAD} SQL function to access the current and next row at the same time. There are options that can be set to allow, disallow or require gaps between ranges.

    {\bf Cross-table checks.}
Ensuring referential integrity between a primary key column in a dataset and a foreign key column in another dataset are low-level functionalities that fall into this category. To the best of our knowledge, no automation or discovery capabilities were found. That means the user has to define themselves the columns to be checked for referential integrity.

    {\bf Value presence / Availability.}
Dbt Core is the only tool of the ones we examined that offers such low-level functionality. The check is as simple as examining whether there is at least one record (value) in the given table (column). The functionality is closely related to the accessibility dimension, since a non-present record or value is obviously inaccessible by the users, in a specific context of use.

    {\bf Number of rows and cardinality.}
The next family of checks relate closely to the completeness dimension of the DQ model. In particular, we encountered two variations of checks about the number of rows in a column or dataset. The first variation verifies constraints on the number of rows, without a reference column or dataset. The user can define either simple or more elaborate size constraints in the form of Spark functions (e.g., \texttt{size > 5 \&\& size < 10}). GX, on the other hand, offers an additional check for the number of \emph{columns} of a dataset.

The second variation measures the number of rows, relatively to another column or dataset. It is essentially a completeness check based on a gold standard. The formal expression is:
\[ \frac{|n|}{|n_{gold}|} \ge T, \]
where $|n|$ is the number of rows in the currently examined column or dataset, $|n_{gold}|$ is the number of rows in the gold standard and $T$ some user-defined threshold in the range $[0, 1]$. Dbt Core supports \texttt{GROUP BY} operations that can be executed before checking completeness. Soda Core can execute checks only for row count equality.
A slightly different completeness check measures the cardinality (i.e. number of distinct values), instead of the total row count. This implementation is, again, based on a gold standard:
\[ \frac{|d|}{|d_{gold}|} \ge T, \]
where $|d|$ is the number of distinct values in the currently examined column, $|d_{gold}|$ is the number of distinct values in the gold standard and $T$ some user-defined threshold in the range $[0, 1]$. Note that cardinality is computed with appropriate SQL queries.

    {\bf Difference between datasets.}
Completeness, however, may not always be perfectly depicted by just the number of rows or distinct values that a column contains. More complex checks may be required, executed both in the available dataset and in the gold standard, in order to assess the results in a comparative way with regard to the reference dataset. MobyDQ and Soda Core offer a more generic, expressive completeness checks. In particular, the user can compute a measure in both the source and target datasets and then get their difference:
\[ \frac{m_t - m_s}{m_s}* 100, \]
where $m_t$ and $m_s$ is the computed value of the metric in the source and target dataset respectively. The result is a percentage representing the difference (in terms of the metric) between the two datasets. Soda core offers similar functionality, with the additional support for more expressive checks of metrics (not only percentage). It also enables the user to define their own query for the source and target datasets in the form of SQL expressions.

    {\bf Matching between datasets.}
A similar low-level functionality  measures the \emph{matching} between a source and a target dataset. In Deequ, the comparison between them can be made in a granularity level of respective columns, defined via a column mapping. The result is the fraction of matching:
\( \frac{|v_{match}|}{|v_{s}|} \ge T, \)
where $|v_{match}|$ is the number of values that match between the two datasets, $|v_{s}|$ is the number of values in the source dataset and $T$ is some user-defined threshold in the range $(0, 1)$. Dbt Core provides the same functionality but supports only full (no partial) matching. Deequ and dbt Core use traditional SQL joins to find the number of matching and mismatching records. Apache Griffin, on the other hand, operates on two Spark dataframes. 

Mapping this family of low-level functionalities to the DQ dimensions can be challenging. More specifically, assuming that the reference dataset plays the role of gold standard, this check is closely related to accuracy and the source dataset is, actually, the source of ground truth. At the same time, counting the number of mismatched records that exist in the source but not in the target implies a completeness check, based, again, on a gold standard. Furthermore, one could argue that currentness is relevant, as well, assuming that data are supposed to flow from source to target. As a result, measuring the fraction of records that appear in the source but not in the target gives an idea about delayed data, imposing freshness issues on the target end. Last but not least, missing data at the target dataset may denote that they are not accessible by the end user (accessibility), since they are not present.

    {\bf NULL values.}
A NULL value denotes an absent value and, thus, affects accuracy, since the true value of the intended attribute fails to be represented. At the same time, a missing value is not available for the end user, thus accessibility is also impaired. Moreover, a dataset with missing values is probably incomplete, making completeness an affected dimension, as well.

From an implementation point of view, different variations of this functionality were found throughout our investigation. Some of them check for total completeness, while others are flexible enough to enable fractional constraints (e.g., \texttt{completeness > 0.9}). The completeness factor can be formally expressed as following:
\[ \mathcal{C} = \frac{|n_{miss}|}{|n|}, \]
where $|n_{miss}|$ denotes the number of missing (NULL) values in the column and $|n|$ the total number of values. Some other variations check completeness on the combination of multiple columns, using either conjuction or disjunction to connect them. Special variations were found for string data. Soda Core and Apache Griffin allow the user to define regex and SQL-like expressions respectively to define  what ``nullness'' means in their own use-case.

    {\bf Distinct \& Unique values.}
Distinct values are considered those that appear \emph{at least} once. For example, in $[a, a, b]$ the distinct values are $a, b$. Unique values are considered those that appear \emph{exactly} once. For example, in $[a, a, b]$ the only unique value is $b$. Unique and distinct values are strongly connected with duplicate data, i.e. values that appear more than once, although they should not. Duplicate values may be deemed as not representing the truth of the intended attribute, so they impair accuracy. The true value is missing, thus it is not available for the user, affecting accessibility as well \cite{classifying_poor_data}.
Measuring distinct and unique values was observed in several variations.

The next big family of checks  involves measuring distinct and unique values in several variations.
Deequ offers three flavors of this metric. The two of them are exact; $\frac{|unique|}{|total|}$ (uniqueness) and $\frac{|unique|}{|distinct|}$ (unique value ratio). Uniqueness can be checked either on single columns or combinations of them. Although there is a reserved primary key check, it offers no extra functionality, since it just checks for uniqueness of the specified column(s). The third flavor offered by Deequ computes an approximation of the distinct values, using HyperLogLog++ sketching. Dbt Core offers built in test only for complete uniqueness, both for single columns and combinations of them. SQL is, again, what tools leverage behind the scenes.

    {\bf Schema-related checks.}
Low-level functionalities in this category include applying constraints on the data types' distribution of a column values, ensuring values in a column are of specific data types and columns exist and are placed in the expected order. Checking for schema-level or data-types discrepancies is strictly connected with both compliance and accessibility dimensions, since parsing wrong data formatting can make data inaccessible to end users. In Deequ, the check's result is a histogram with the data types distribution. It allows filtering both by relative and absolute fractions. Soda Core implements schema-level checks, including presence, absence or order of a column in a dataset. It can also be checked whether a column has the expected data type. An interesting implementation detail is the fact that the column names defined by the user can contain wildcards, such as \% or $\star$, depending on the wildcards that are supported in each data source. GX provides a dedicated check for date data.

    {\bf Recentness.}
Lastly, we subsume here various low-level functionalities that are directly related to the currentness dimension. In particular, the checks operate on a column that contains timestamps and tries to enforce that all values in the column are at least as recent as a given time interval:
\( \forall i \in [0, N), \text{\hspace{1 mm}} t_i + t_{interval} \ge t_{current}, \)
where $N$ is the total number of values in the timestamp column, $t_i$ is the timestamp value at index $i$, $t_{interval}$ is the user-defined timeliness interval and $t_{current}$ is the current system time.

Different variations of this functionality were found. The main difference among them has to do with the way the reference timestamp is considered. When current timestamp is considered, we talk about ``freshness'' of data. When the latest timestamp of a source dataset is used, we talk about ``latency'', depicting how quickly data flow from target to source, regardless of current timestamp.

\section{Related work\label{section:related_work}}
    DQ management has been an active research area. Ehrlinger \& Wo{\ss}~\cite{survey_ehringer} have conducted a large-scaled systematic survey on 13 DQ tools, including both open-source and commercial ones. Among other contributions, they provide insights regarding the DQ metrics and dimensions supported by the tools they examined. There are significant differences in our approaches, since they highlight data profiling and automated DQ monitoring as focal points of their selection (i.e., inclusion and exclusion) criteria. On the other hand, we do not impose such strict criteria, resulting to different and newer tools. Among the 6 tools examined in the current survey and the 13 tools discussed in~\cite{survey_ehringer}, only two of them (MobyDQ and Apache Griffin) are in common. The main point our approach differentiates is the thorough gathering, grouping and mapping of \emph{low-level} DQ functionalities to DQ dimensions, focusing more on engineering aspects, closer to the programming implementation, while the authors in \cite{survey_ehringer} opt for a detailed, yet more abstract classification of functionalities.
    Goknin et al.~\cite{DQ_iot_industry_4_0} study the problem of DQ measurement from the scope of CPS and IoT (Industry 4.0), providing insights, taxonomies and paradigms about various aspects of how DQ mechanisms are adapted into industrial settings. Laranjeiro et al.~\cite{classifying_poor_data} on the other hand provide a mapping, with the difference that they start from DQ problems towards dimensions, while we opt for starting from DQ solutions being used in real-life scenarios. The main reason for this choice has been to attempt to capture the materialization of DQ dimensions in practice.

\section{Conclusion and future work\label{section:conclusion_future_work}}
    In this report, we have investigated six open-source DQ tools in terms of the low-level functionalities they offer. We comment on the implementation of these functionalities, mapping them with the higher-level DQ dimensions they are closely related to. Regarding future steps, more research effort should be devoted to identify the connection between theoretical models of DQ and how challenges are solved in practice with a view to extracting quantifiable DQ metrics in a more systematic way.

    Bypassing the discrepancies in terminology and standardizing the DQ management implementation aspects would be of great benefit in designing DQ solutions that address modern challenges effectively. Last but not least, the vast majority of open-source tools to date do not deal with streaming data, and thus, adapting DQ solutions to online settings should attract more attention.

\bibliographystyle{ACM-Reference-Format}
\bibliography{sample}


\begin{thebibliography}{8}


\ifx \showCODEN    \undefined \def \showCODEN     #1{\unskip}     \fi
\ifx \showDOI      \undefined \def \showDOI       #1{#1}\fi
\ifx \showISBNx    \undefined \def \showISBNx     #1{\unskip}     \fi
\ifx \showISBNxiii \undefined \def \showISBNxiii  #1{\unskip}     \fi
\ifx \showISSN     \undefined \def \showISSN      #1{\unskip}     \fi
\ifx \showLCCN     \undefined \def \showLCCN      #1{\unskip}     \fi
\ifx \shownote     \undefined \def \shownote      #1{#1}          \fi
\ifx \showarticletitle \undefined \def \showarticletitle #1{#1}   \fi
\ifx \showURL      \undefined \def \showURL       {\relax}        \fi
\providecommand\bibfield[2]{#2}
\providecommand\bibinfo[2]{#2}
\providecommand\natexlab[1]{#1}
\providecommand\showeprint[2][]{arXiv:#2}

\bibitem[gri(2024)]%
        {griffin_website}
 \bibinfo{year}{2024}\natexlab{}.
\newblock \bibinfo{title}{Apache Griffin}.
\newblock
\newblock
\urldef\tempurl%
\url{https://griffin.apache.org/}
\showURL{%
\tempurl}


\bibitem[app(2024)]%
        {approximate_quantile_deequ}
 \bibinfo{year}{2024}\natexlab{}.
\newblock \bibinfo{title}{Deequ's Approximate Quantiles}.
\newblock
\newblock
\urldef\tempurl%
\url{https://github.com/awslabs/deequ/blob/master/src/main/scala/ com/amazon/deequ/analyzers/ApproxQuantile.scala\#L50}
\showURL{%
\tempurl}


\bibitem[iso(2024)]%
        {iso25012}
 \bibinfo{year}{2024}\natexlab{}.
\newblock \bibinfo{title}{ISO/IEC 25012}.
\newblock
\newblock
\urldef\tempurl%
\url{https://iso25000.com/index.php/en/iso-25000-standards/iso-25012}
\showURL{%
\tempurl}


\bibitem[atl(2024)]%
        {atlan_list}
 \bibinfo{year}{2024}\natexlab{}.
\newblock \bibinfo{title}{Open-source data quality tools}.
\newblock
\newblock
\urldef\tempurl%
\url{https://atlan.com/open-source-data-quality-tools/}
\showURL{%
\tempurl}


\bibitem[Ehrlinger and Wöß(2022)]%
        {survey_ehringer}
\bibfield{author}{\bibinfo{person}{Lisa Ehrlinger} {and} \bibinfo{person}{Wolfram Wöß}.} \bibinfo{year}{2022}\natexlab{}.
\newblock \showarticletitle{A Survey of Data Quality Measurement and Monitoring Tools}.
\newblock \bibinfo{journal}{\emph{Frontiers in Big Data}}  \bibinfo{volume}{5} (\bibinfo{year}{2022}).
\newblock
\showISSN{2624-909X}
\urldef\tempurl%
\url{https://doi.org/10.3389/fdata.2022.850611}
\showDOI{\tempurl}


\bibitem[Goknil et~al\mbox{.}(2023)]%
        {DQ_iot_industry_4_0}
\bibfield{author}{\bibinfo{person}{Arda Goknil}, \bibinfo{person}{Phu Nguyen}, \bibinfo{person}{Sagar Sen}, \bibinfo{person}{Dimitra Politaki}, \bibinfo{person}{Harris Niavis}, \bibinfo{person}{Karl~John Pedersen}, \bibinfo{person}{Abdillah Suyuthi}, \bibinfo{person}{Abhilash Anand}, {and} \bibinfo{person}{Amina Ziegenbein}.} \bibinfo{year}{2023}\natexlab{}.
\newblock \showarticletitle{A Systematic Review of Data Quality in CPS and IoT for Industry 4.0}.
\newblock \bibinfo{journal}{\emph{ACM Comput. Surv.}} \bibinfo{volume}{55}, \bibinfo{number}{14s}, Article \bibinfo{articleno}{327} (\bibinfo{date}{jul} \bibinfo{year}{2023}), \bibinfo{numpages}{38}~pages.
\newblock
\showISSN{0360-0300}
\urldef\tempurl%
\url{https://doi.org/10.1145/3593043}
\showDOI{\tempurl}


\bibitem[Karnin et~al\mbox{.}(2016)]%
        {kll_sketch}
\bibfield{author}{\bibinfo{person}{Zohar Karnin}, \bibinfo{person}{Kevin Lang}, {and} \bibinfo{person}{Edo Liberty}.} \bibinfo{year}{2016}\natexlab{}.
\newblock \showarticletitle{Optimal Quantile Approximation in Streams}. In \bibinfo{booktitle}{\emph{2016 IEEE 57th Annual Symposium on Foundations of Computer Science (FOCS)}}. \bibinfo{pages}{71--78}.
\newblock
\urldef\tempurl%
\url{https://doi.org/10.1109/FOCS.2016.17}
\showDOI{\tempurl}


\bibitem[Laranjeiro et~al\mbox{.}(2015)]%
        {classifying_poor_data}
\bibfield{author}{\bibinfo{person}{Nuno Laranjeiro}, \bibinfo{person}{Seyma~Nur Soydemir}, {and} \bibinfo{person}{Jorge Bernardino}.} \bibinfo{year}{2015}\natexlab{}.
\newblock \showarticletitle{A Survey on Data Quality: Classifying Poor Data}. In \bibinfo{booktitle}{\emph{2015 IEEE 21st Pacific Rim International Symposium on Dependable Computing (PRDC)}}. \bibinfo{pages}{179--188}.
\newblock
\urldef\tempurl%
\url{https://doi.org/10.1109/PRDC.2015.41}
\showDOI{\tempurl}


\end{thebibliography}

\appendix
\newpage
\begin{table*}[!ht]
    \scriptsize
    \centering
    \begin{center}
        \bgroup
        \def\arraystretch{1.5}
\begin{tabular}{|p{2in}|p{4.8in}|}
            \hline
            \multicolumn{1}{|c|}{\textbf{Low-level functionality}}       & \multicolumn{1}{c|}{\textbf{Function name(s)}} \\ \hline                                                                                                                 
            anomaly detection based on comparison with saved statistics                 & Deequ (\texttt{isNewestPointNonAnomalous})                                                                                                                                                                                                                                                                                                                                                                                                                                                                                                                                                                                                                                                                                                                             \\ \hline 
            max / min / mean / median / std / sum / z-score of values of a column       & Deequ (\texttt{hasMin}, \texttt{hasMax}, \texttt{hasMean}, \texttt{hasSum}, \texttt{hasStandardDeviation}), Soda Core (\texttt{avg}, \texttt{max}, \texttt{min}, \texttt{sum}, \texttt{stdev}, \texttt{stdev\_pop}, \texttt{stdev\_samp}, \texttt{variance}, \texttt{var\_pop}, \texttt{var\_samp}), GX (\texttt{expect\_column\_max\_to\_be\_between}, \texttt{expect\_column\_mean\_to\_be\_between}, \texttt{expect\_column\_median\_to\_be\_between}, \texttt{expect\_column\_min\_to\_be\_between}, \texttt{expect\_column\_stdev\_to\_be\_between}, \texttt{expect\_column\_sum\_to\_be\_between}, \texttt{expect\_column\_value\_z\_scores\_to\_be\_less\_than}, \texttt{expect\_column\_values\_to\_be\_between}, \texttt{expect\_multicolumn\_sum\_to\_equal}) \\
\hline 
        \end{tabular}
        \egroup
    \end{center}
    \caption{Appendix: The exact function names employed by the tools that implement the studied DQ low-level functionalities.}
    \label{table_functionalities_names}
\end{table*}

\end{document}